# "LA NUBE": LA MANERA MÁS EMOCIONANTE DE EXPERIMENTAR ESPEJOS PLANOS

José J. Lunazzi
Instituto de Física - Universidad Estatal de Campinas – UNICAMP
Campinas – SP-BR     lunazzi@ifi.unicamp.br

**Resumen**
Una experiencia que parte del uso de un espejo plano colocado horizontalmente a la altura de la nariz del observador, que permite uma nueva dimensión a la experiência de ver imágenes com espejos simples, sirve para discutir lo que aún puede hacerse para que el estudiante gane uma visión más profunda y conceptual en el aprendizaje de la física. El énfasis está en la idea de independéncia mental, que incluye la de que nuevos resultados pueden surgir en el análisis de experimentos que no necesáriamente precisan ser hechos con elementos nuevos o de sofisticada realización.

Resumo
Uma experiência que parte do uso de um espelho plano colocado horizontalmente na altura do nariz do observador, que fornece uma nova dimensão à experiência de se ver imagens com espelhos simples, serve para discutir o que ainda pode ser feito para que o estudante ganhe uma visão mais profunda e conceitual no aprendizado de física. A ênfase está na idéia de independência mental, que inclui a de que novos resultados podem surgir na análise de experimentos que não necessariamente precisam ser feitos com elementos novos ou de sofisticada realização.

Resumo
Eksperimento bazita el la horizontala uzado de ebena spegulo lau pozicio de la nazo de observanto donas nova dimensio al la eksperto vidi bildoj per simplaj speguloj kaj utilas por diskuti kio eblas ankorau fari por studento akiri plej profonda kaj koncepta pensmaniero dum lernado de fiziko. La kerno estas la ideo de mensa sendependeco, kio inkluzivas la eblecon novaj rezultoj esti atingitaj dum analizo de eksperimentoj kiuj ne nepre bezonas esti faritaj per novaj au tre specialaj elementoj.

Abstract
The experience of employing a plane mirror located at the height of the nose of an observer, giving a new dimension to the experiences made with simple mirrors, is used as an example into the discussion of what can be done for the student to reach a deeper and conceptual insight in learning Physics. The main idea lays on the mental independence, which includes that new results can come through the analysis of experiments which not necessarily needs to be made with new elements or through a sophisticated procedure.

## I. Introducción

Una pregunta fué formulada en la conferencia dada por un Prémio Nobel de Física estadounidense en Brasil [1]: "Si usamos en la universidad los mismos libros de texto que ustedes, ¿por que no tenemos prémios Nobel?" Permitanmé una conjetura, que el lector podrá comparar con las suyas: es que no se trata de colocar nominalmente un libro de texto, sinó de las horas dedicadas a clases específicamente de Física y de la profundidade de su tratamiento. La demonstración de las leyes y de los princípios teóricos correspondientes, en general, no se exige aquí. Considero que, partindo del teorema de Pitágoras, del princípio de Arquímedes, en fin, siguiendo la evolución de la ciencia, cualquiera de nosotros, físicos, deberia poder probar la validad de las leyes que utiliza. Aceptarlas simplemente y aplicarlas en ejercícios típicos ya constituye una falta de conocimiento y de posibilidades de repensar los fundamentos o los puntos no tratados y las condiciones establecidas en esas leyes. He visto ocurrir, con textos como los de la serie de Resnick-Haliday, que muchos profesores al terminar la presentación del texto de un capítulo, van directamente a los ejercícios eliminando las preguntas que existen al final de éste. Se elimina, así, la posibilidad de una amplitud mayor de la mente del estudiante, que vá a quedar limitado a los casos de los ejercícios.

Y, finalmente, algo también primario: nuestras clases teóricas de los cursos básicos no suelen ser realizadas, como en el primer mundo, con el profesor delante de una extensa mesa llena con experimentos que inducen a las fórmulas y a las conclusiones teóricas. No es que la Física no pueda ser estudiada unicamente desde el punto de vista teórico, mas, si nos limitamos a las fórmulas, sim una visión de lo que el experimento es materialmente, perdemos chances de entrar en el Nobel.

Otro factor a destacar es el de la dependencia ideológica. Mientras que en los EUA la independencia política con el colonizador fué antes que en el resto de América y de manera bien definida por uma guerra, en Brasíl fué casi um entendimiento, una negociación. Después de la salida del Imperio Portugués, la fuerte presencia de Inglaterra, inicialmente, y de los EUA, después, hizo mantener el peso imperial sobre el ideario de las personas. "Otros son más ricos, más fuertes y más inteligentes que nosotros, hablan una lengua que no es facil, que nos cuesta entender, las autoridades los tratan mejor que a nosotros mismos", podria pensar el ciudadano común, y también nuestro alumno. Sin considerar aún el hecho de que la mayoría de la población queda marcada con los prejuicios subyacentes de la esclavitud, por la inferioridad económica y por el color de su piel. Resumiendo, la idea de colonia no combina con la de Premio Nobel. Como en el deporte, es preciso que el aspirante crea en sus posibilidades. Falta enseñar a valorizar lo que los países latinos de América han creado, incluyendo sua historia y prehistoria. Este trabajo muestra una manera, aunque modesta, de hacer eso.

¿Como llegar al Nobel? Acabo de leer un dibujo humorístico argentino donde un pintor se enfrenta con una tela y con su pincel listo para comenzar a pintar un cuadro se pregunta: "¿Cómo se hace para pintar un cuadro de sesenta y cinco milllones de dólares?". Nadie enseña, claro, pero una respuesta podria ser: yendo a lo simple, a lo básico, a la eséncia del problema, entendiendo lo que no encaja bien e intentando una solución. Agregar detalles a lo que existe sería algo relativamente convencional. Cuando vemos el trabajo que llevó a um premio Nobel siendo explicado con simplicidad, pensamos: ¿Mas por que nadie pensó antes en eso? Está ahí el centro de la cuestión. El multimilionario griego Onassis explicaba su fortuna a un periodista diciendo: "¿Ve esa lámpara?" (un objeto atrás del periodista). El se dá vuelta y responde: "Sí". "Pues yo la ví primero", dijo Onassis, terminando su explicación. Aún correspondiendo al comercio o al arte, ejemplos de iniciativa pueden ser válidos para nuestra propuesta.

## II. Descubriendo los espejos planos

Daré un ejemplo de que siempre debemos estar atentos para descubrimientos aún en los asuntos más simples y conocidos. Las referéncias más antiguas que conozco del uso de espejos corresponden a más de dos mil años de antigüedad y a tres regiones geográficas: América Latina, China y Grecia. En la primera, hecho prácticamente desconocido por el público y aún por los físicos, se tienen espejos con diferentes radios de curvatura, convergentes y divergentes obtenidos puliendo rocas. Existen también monumentos e iconografía que muestran la realización de imágenes con esos espejos en territorio que hoy constituye Méjico [2]. También fueron encontrados espejos planos en Perú[3]. Todo indica que la primacía de los espejos de mayor calidad, los que forman imágenes, corresponde a regiones próximas al Brasil y a personas que hablaban lenguas que no eran europeas. La estética del brillo y el color ciertamente han estimulado eso [4].

En el caso de Grecia, parecería que los espejos, siendo metálicos, no podian tener buen pulimiento y solamente habrían dado imágenes de poca calidad o no quedaron bien preservados [5], o tal vez sirvieron para quemar embarcaciones enemigas [6]. Y menos aún podríamos citar imágenes de reflexión en China, donde, por lo que se sabe, los espejos servirían apenas para concentrar el sol para hacer fuego [7], para cicatrización en medicina, o por su valor decorativo [8]. Contemos eso a nuestros alunos para que entiendan que no hay nada de extraordinario en el escenario que puede llevar a un desarrollo científico, será siempre la inteligencia el factor principal, y no factores geopolíticos.

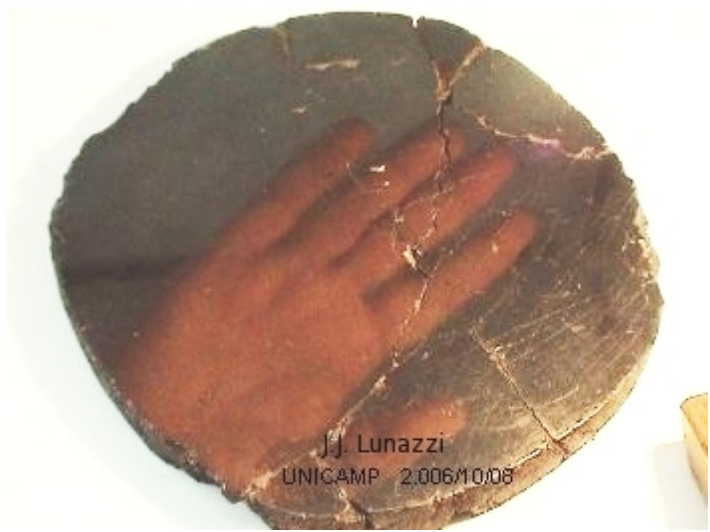

*Fig. 1  Foto de un espejo plano precolombino reflejando la mano del autor.*

¿Sería posible hacer algo nuevo con los espejos planos simples, de esos que usamos en casa para ver nuestro rostro, o como elemento decorativo en los corredores y espacios de tránsito y recepción de visitas? Desde pequeños, billones de personas ya experimentaron el manoseo de ellos. Han sido usados para experimentos didácticos durante siglos. La descripción teórica que demuestra que la imagen ocurre en un espacio virtual simétrico al del objeto hace parte del conocimiento obligatorio de todo alumno de escuela secundaria. Y, sin embargo, hace menos de treinta años que Mireya Baglietto, una artista argentina [9], tuvo la idea de colocar un espejo horizontalmente a la altura de la nariz para representar al mundo cabeza abajo. Ver abajo lo que está encima nuestro.

Esa subversión del espacio visual, a pesar de no alterar el fenómeno físico, es sorprendente y confunde al sentido del equilíbrio. El comando de los pasos resulta mu afectado y la persona anda lentamente y con cuidado. Baglietto ha creado espacios específicos en forma de tiendas con túneles donde hay elementos colgados cuyas imágenes surgen delante de las personas cuando los recorren. Al avanzar, la persona se vé atravesando mágicamente los objetos, inmateriales. Se pone al público de modo que sus sapatos pisen siempre sobre material blando, o idealmente el piso lo es, para quitarle firmeza también al sentido del tacto en el apoyo. Un sonido electrônico genera sensaciones indefinidas, todo para liberar de condiciones a la percepción de la persona. El uso artístico, y hasta terapéutico, viene siendo explorado por ella hasta hoy.

Pensamos, hce dos años, en hacer uso didático y hemos usado espacios que se encuentran en cualquier corredor o patio, sean estos cerrados o abiertos.  La idea es entusiasmar por medio del uso de la óptica, hacer al público vivenciar lo que es una imagen simétrica, que puede ocurrir, ahora, también en la dirección vertical. La simetria en el sentido horizontal ni siquiera es notada por la mayoria das personas en el uso diario, donde los espejos están colocados verticalmente. Pero esta simetría nuestra sí, resulta inolvidable [10]. En un espacio cerrado, el techo se vé bien abajo de donde se espera ver el piso, la sensación de sorpresa es fuerte y la inestabilidad de nuestra posición nos hace dudar de nuestra seguridad. Podemos colocar objetos a media altura, colgados del techo, inmediatamente sobre la cabeza de las personas, para que les aparezcan delante y puedan atravesarlos. Hemos puesto espadas falsas para aumentar el dramatismo. La persona duda si debe ir adelante, metiéndose en la imagen, que parece el objeto verdadero.

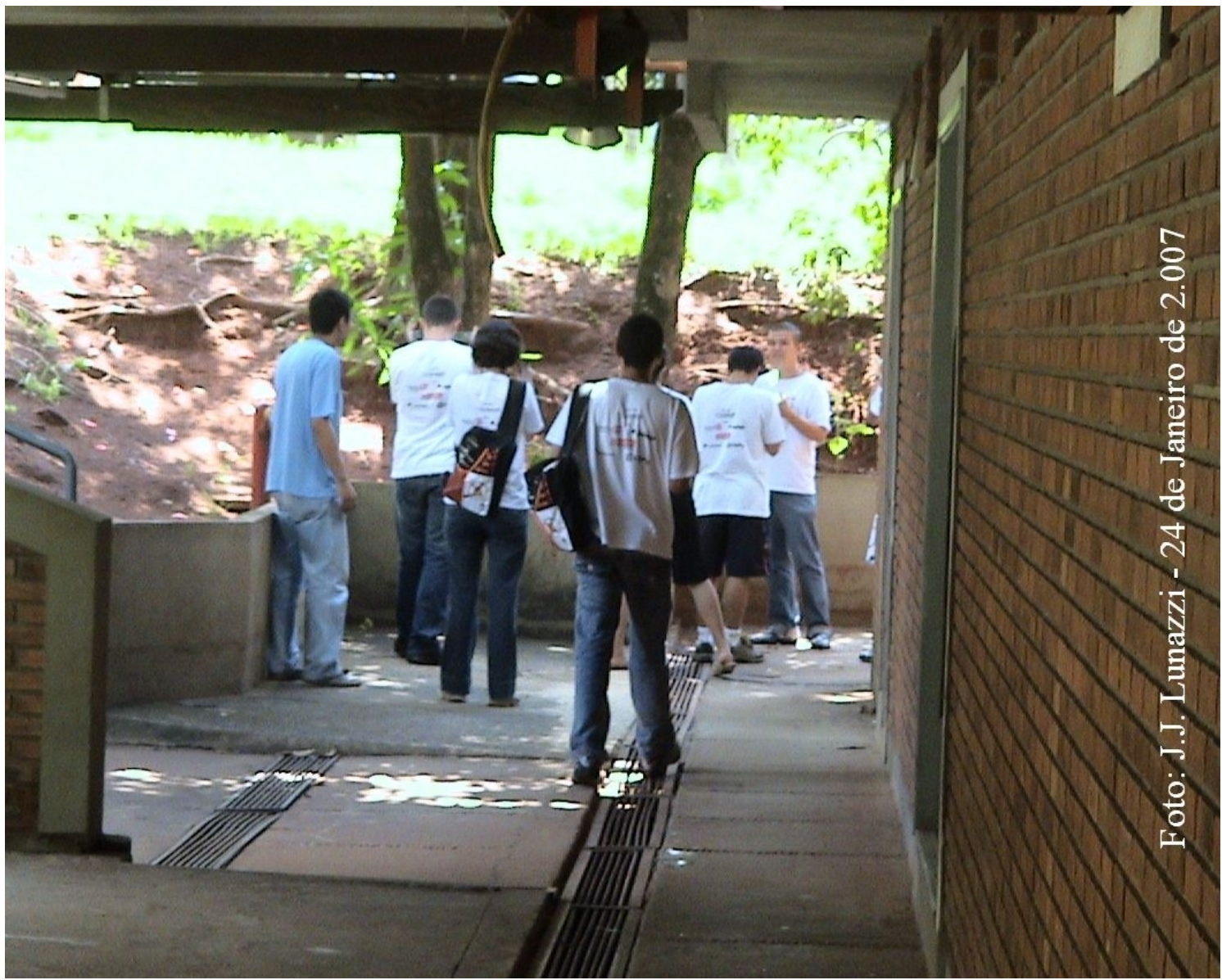

*Fig. 2  Alumnos participando de la experiéncia.*

En un espacio abierto tenemos el cielo y, habiendo nubes, por el espejo nos vemos caminando muy sobre ellas como solamente desde um avión sería possible. No conseguimos la misma sensación si nos acostamos y miramos para arriba: sabemos que en ese caso lo que vemos ocurre encima. El nombre artístico de la experiência: "La Nube" evoca la sensación de flotar en el aire. Llegando cerca de árboles vemos los troncos como columnas que nos conducen al vacío, allá abajo, los troncos siendo la única esperanza de sobrevivencia porque podríamos agarrarnos a ellos para no caer. Lo mejor, entonces, es hallar un espacio cubierto con salida para uno abierto, como vemos en la Fig. 2. El vacío aparece de repente, sorprende. Habrían tal vez los antíguos usado los espejos de esa manera? Podemos dudar de ello, porque no conocemos representaciones de personas con espejos así colocados. El poder reflectivo de los espejos precolombinos es poco, por otra parte, y el tamaño algo pequeño, 15 cm x 20 cm, para pensar en esa posibilidad. Sin embargo, se conocen estatuetas donde personas llevam una cuba chata en esa posición, sugiriéndo que estuviese llena con agua y por lo tanto reflejase. La presencia de uma cruz del tipo de Malta dibujada en la cabeza de la persona ha hecho pensar, sin embargo, en otro fenómeno: la visualización de un efecto de polarización, pues con fondo obscuro y ángulo próximo al de Brewster la polarización de la luz del cielo podria tener el efecto de un analizador [11], generando una cruz así.

### III. Detalhes experimentales

El espejo debe trener 24 cm x 30 cm x 0,2 cm. Un espesor de 6 mm genera uma estabilidade maior, lo que es bueno, pero el costo es cerca de tres veces mayor y el peso podria ser sentido como la presencia de un dispositivo, lo que es indeseable. El corte para encaje de la nariz sigue o modelo de la Fig. 3 [12] y puede ser encargado a una vidrieria. No es indispensable y trae el riesgo de que una imperfección del corte se propague estropeando la pieza, pero aumenta el campo próximo de visión e indica directamente como debe ser usado. Evita que el espejo sea colocado abajo de la nariz, con lo que la imagen de la nariz podria estar presente y la humedad de la respiración también.

El modelo de la figura muestra un borde de goma protegiendo contra golpes, e incluye una capa de goma abajo. Posteriormente, este fué substituido por uno que utiliza solamente uma capa de goma abajo del espejo, sobresaliendo de 1 a 2 cm. Esto tiene dos ventajas: elimina la separaçãon visual entre el campo del espejo y el de la visión exterior a él, evitando la idea de un mundo en el espejo y otro afuera. Al mismo tiempo, evita el despegamiento del borde de goma opr el manoseo. Cuanto exactamente la goma puede sobresalir del espejo deberá ser estimado visualmente, pues depende del espesor elegido para el espejo.

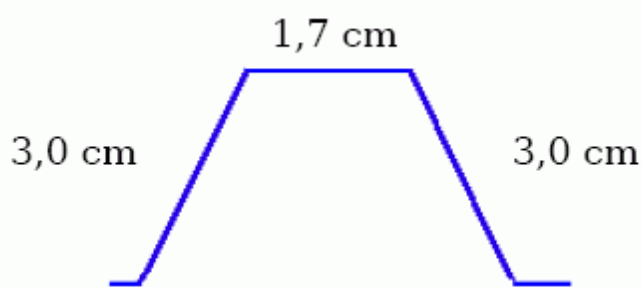 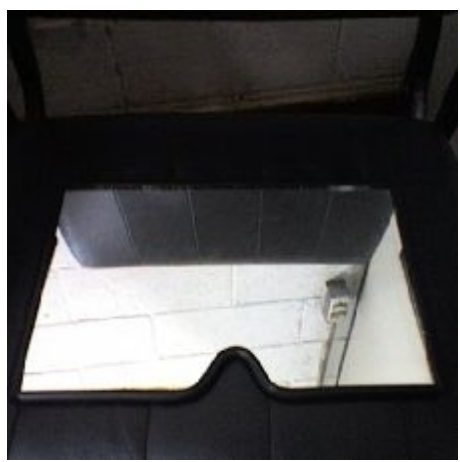

*Fig. 3  Modelo de espejo. Izquierda: datos para el recorte. Derecha: foto.*

### IV. Medidas de seguridad

El único inconveniente notado viene de la posibilidade de un accidente, aunque esto nunca haya ocurrido apesar de que más de novecientas personas já experimentaron el fenómeno. Él podria ser mecánico, con la persona cayendo y el espejo quebrandosé de manera cortante. Un espejo de acrílico pareciera ser la solución, mas el costo sería cuatro veces mayor y la conservación dificílisima: sería casi imposible conseguir que nadie tocase la superficie óptica, y a limpieza y el roce del almacenamiento causarían rayas, afectando la calidad al punto de depreciar mucho el resultado. Existe acrílico que puede ser limpiado, que usamos en los anteojos y vemos
en las ventanas de avión, mas este no es el que se compra en comércios convencionales. Intentamos hallar un material transparente que cubriese la superficie, pero ni los vendidos en papelerías para cubrir cuadernos y libros, ni el material para conservación de alimentos, ni los existentes para protección de pantalla de cristal líquido para computadores consiguen manteener la calidad de la imagen. Es que tienen un grdo de distorción los primeiros, y de difusión los últimos, por los que una simple rugosidad o su grado de difusión antirreflejo provoca una distorción en la imagen. La única opción contra accidentes es tomar mucho cuidado con la presencia de desniveles en el piso, advirtiendo a las personas para andar despacio, deslizando los piés, y observando la conducta de ellas para controlar personas inquietas, tales como niños. Una capa de goma tipo espuma puede ser colada debajo del espejo para amortiguar un eventual impacto en caso de caida.

El accidente podria também ser óptico: en algun momento del recorrido un rayo de Sol puede incidir en los ojos y sabemos que eso puede quemar puntos de la retina. Para eso, es necesário un análisis previo del recorrido, indicando cuando necesario que el camino a seguir es aquel que deja al Sol a espaldas de la persona, no utilizando el espejo al regresar. El Sol exactamente encima del público anula el experimento. Se debe estar atento también a cambios de posición de este, probando el recorrido de hora em hora (es conveniente fijar estas recomendaciones en el dorso del espejo).

 *******************************************************************************